# High atmospheric metal enrichment for a Saturn-mass planet

Jacob L. Bean[1], Qiao Xue[1,2], Prune C. August[1,3], Jonathan Lunine[4], Michael Zhang[1], Daniel Thorngren[5], Shang-Min Tsai[6,7], Keivan G. Stassun[8], Everett Schlawin[9], Eva-Maria Ahrer[10], Jegug Ih[11], & Megan Mansfield[9]

[1] Department of Astronomy & Astrophysics, University of Chicago, Chicago, IL, USA
[2] School of Physics and Astronomy, Shanghai Jiaotong University, Shanghai, China
[3] Institute of Physics, Ecole Polytechnique Fédérale de Lausanne, Lausanne, Switzerland
[4] Department of Astronomy, Cornell University, Ithaca, NY, USA
[5] Department of Physics & Astronomy, Johns Hopkins University, Baltimore, MD, USA
[6] Department of Earth Sciences, University of California, Riverside, California, USA
[7] Department of Physics, University of Oxford, Oxford, UK
[8] Department of Physics & Astronomy, Vanderbilt University, Nashville, TN, USA
[9] Steward Observatory, University of Arizona, Tucson, AZ, USA
[10] Department of Physics, University of Warwick, Coventry, UK
[11] Department of Astronomy, University of Maryland, College Park, MD, USA

**Atmospheric metal enrichment (i.e., elements heavier than helium, also called "metallicity") is a key diagnostic of the formation of giant planets[1-3]. The giant planets of the solar system exhibit an inverse relationship between mass and both their bulk metallicities and atmospheric metallicities. Extrasolar giant planets also display an inverse relationship between mass and bulk metallicity[4]. However, there is significant scatter in the relationship and it is not known how atmospheric metallicity correlates with either planet mass or bulk metallicity. Here we show that the Saturn-mass exoplanet HD 149026b[5-9] has an atmospheric metallicity 59 - 276 times solar (at 1σ), which is greater than Saturn's atmospheric metallicity of ~7.5 times solar[10] at >4σ confidence. This result is based on modeling $CO_2$ and $H_2O$ absorption features in the thermal emission spectrum of the planet measured by JWST. HD 149026b is the most metal-rich giant planet known, with an estimated bulk heavy element abundance of 66±2% by mass[11,12]. We find that the atmospheric metallicities of both HD 149026b and the solar system giant planets are more correlated with bulk metallicity than planet mass.**

We measured the dayside thermal emission spectrum of HD 149026b using the NIRCam instrument[13] on JWST to investigate the composition of the planet's atmosphere in the context of its high bulk metallicity. The observations consisted of two visits of 8.27 hours each centered on secondary eclipses of the planet. Both observations used the module A grism R mode to obtain time-series near-infrared spectra in the long-wavelength channel. The first observation on July 15, 2022 used the F322W2 filter to obtain data from 2.349 to 4.055 μm, while the second observation on August 4, 2022 used the F444W filter to obtain data from 3.778 to 5.082 μm. The observations both used the SUBGRISM64 subarray and the BRIGHT2 readout pattern. The first observation obtained 5 groups per integration, yielding 7935 total integrations. The second observation obtained 9 groups per integration, yielding 4597 total integrations. De-focused, short-wavelength filter photometry was also obtained simultaneously during both visits using the WLP4 weak lens. This lens is joined with a F212N2 narrowband filter that is centered at 2.12 μm and has a 2.3% bandpass.

We reduced and analyzed the JWST data using the Eureka! pipeline[15]. We followed similar steps as those outlined in the recent Transiting Exoplanet Community Early Release Science Team paper on NIRCam transits of the planet WASP-39b[16]. We validated our implementation of the Eureka! pipeline on the NIRCam commissioning time-series data for the planet HAT-P-14b[17]. We summed the resulting extracted spectra to create light curves that we then modeled to measure the planet's thermal emission. We created both "white" light curves that were summed over the full bandpass of each observation, and spectroscopic light curves that were summed over 20 pixels each (yielding 140 channels in total). We also performed aperture photometry on the short-wavelength filter photometry to produce additional time-series data. However, the short-wavelength photometry from the second visit exhibits significant correlated noise from an unknown source that does not impact the long-wavelength spectroscopy. We ultimately excluded both short-wavelength data sets from our subsequent analysis because of this additional noise; their omission does not affect the results because they do not add significant additional information compared to the spectra. See the Methods for more details.

We modeled the white and spectroscopic light curves from the long-wavelength data using a combination of models for the planet's thermal emission and the systematics in the data. Both data sets exhibit a quick, exponential-like ramp downward over approximately the first hour of the data, followed by a near-linear drift in time (see Extended Data Figure 5). The weak lens photometry indicates that the first data set exhibits a mirror segment "tilt event"[17] right before the secondary eclipse. However, this event did not have an obvious impact on the long-wavelength spectroscopy and so we didn't perform any corrections for it. The data also exhibit at most a weak dependence on spectral position in the cross dispersion direction, and no measurable dependence on the position in the dispersion direction. In our primary analysis we trimmed the first 30 minutes of data from the light curves and then modeled the remaining drift as the product of exponential and linear functions with respect to time. We obtained similar results when not trimming the first 30 minutes of data and when including a linear decorrelation with position in the cross dispersion direction. We assumed that the star plus planet flux is uniform out of the eclipse because the phase curve signal of the planet is negligible over our short observational baseline[18].

Our model for the secondary eclipse assumed the fixed parameters planet orbital period P = 2.87588874 days, ratio of the planet's semi-major axis to the host star radius $a/R_s$ = 6.36, planetary orbital inclination = 86.7°, and planetary orbital eccentricity e = 0.0; varying these parameters within their uncertainties did not impact the results. We determined the secondary eclipse times for each observation from fitting the white light curves and then fixed them in the analysis of the spectroscopic light curves. The final fits solve for the planet-to-star flux ratio ($F_p/F_s$) and the systematics model parameters. The dayside thermal emission spectrum is constructed from the $F_p/F_s$ values as a function of wavelength. The rms of the residuals to the spectroscopic light curves are mostly within a few tens of percent of the photon noise, while a total of 14 of the 140 spectroscopic channels have residuals that exceed 1.4 times the photon noise. The results we report below are unchanged if we exclude the points with large errors. The six reddest and six bluest data points from the first and second visits, respectively, overlap in wavelength from 3.889 - 4.009 μm. The visit-level weighted means of the eclipse depths for these overlapping data points agree to better than 2 ppm.

Our JWST dayside thermal emission spectrum of HD 149026b is shown in Figure 1. The previous broadband photometry for this planet from the Spitzer Space Telescope were suggestive of the presence of strong CO and/or $CO_2$ absorption features around 4.5 μm, which would be expected if the planet has a super-solar metallicity atmosphere[14,18]. The JWST data are consistent with three of the four previous Spitzer photometric points at better than 2σ when integrating over their respective bandpasses. The 3.6 μm Spitzer point of Ref. [18] is inconsistent with the JWST data at 4.4σ, but this channel is known to have worse data quality than the 4.5 μm Spitzer channel. Our results agree better with the 3.6 μm Spitzer point of Ref. [14]. Ultimately, the JWST data spectroscopically resolve HD 149026b's thermal emission for the first time. The spectrum deviates from a blackbody at 9.2σ confidence. The best-fit brightness temperature is 1,784 K, which can be compared to the planet's zero-albedo equilibrium temperature of 1,706 K. We conclusively identify (at 3.2σ confidence) the same $CO_2$ band at 4.2 μm that was recently reported in the transmission spectrum of WASP-39b[19]. Modeling the data assuming thermochemical equilibrium (see below) suggests that the continuum across most of the bandpass is set by $H_2O$ absorption.

We performed atmospheric retrieval on the JWST spectrum of HD 149026b using the PLATON code[20,21]. This 1D code assumes scaled solar abundances, thermochemical equilibrium, and a parameterized version of the temperature-pressure (T-P) profile from Ref. [22]. We implemented the dilution factor advocated by Ref. [23] to account for inhomogeneities across the dayside hemisphere that could bias our 1D retrieval. In total, the retrieval has eight free parameters: the logarithm of the metallicity relative to solar ([M/H]), the carbon-to-oxygen (C/O) ratio, five T-P profile parameters, and the dilution factor. Figure 1 shows the best fit to the data. This model has a reduced $\chi^2$ = 0.94. The modeling reinforces the identification of $CO_2$ and $H_2O$ and demonstrates that other molecules that absorb in the bandpass are not detected. The three low points in the spectrum near 4.1 µm are suggestive of the $SO_2$ feature that was also recently detected in WASP-39b[24,25]. We performed photochemistry calculations following the methods of Ref. [26] and attempted to constrain the $SO_2$ abundance using the data. However, our results only yielded a 2σ upper limit on the logarithm of the $SO_2$ volume mixing ratio of -4.1, which is about an order of magnitude larger than the abundance expected from photochemistry in a 100x solar metallicity atmosphere. Given this result and the excellent fit quality of our standard (i.e., only considering chemical equilibrium) model, we conclude that $SO_2$ is not detectable in our data and also that this non detection is not inconsistent with the expectations from photochemistry calculations.

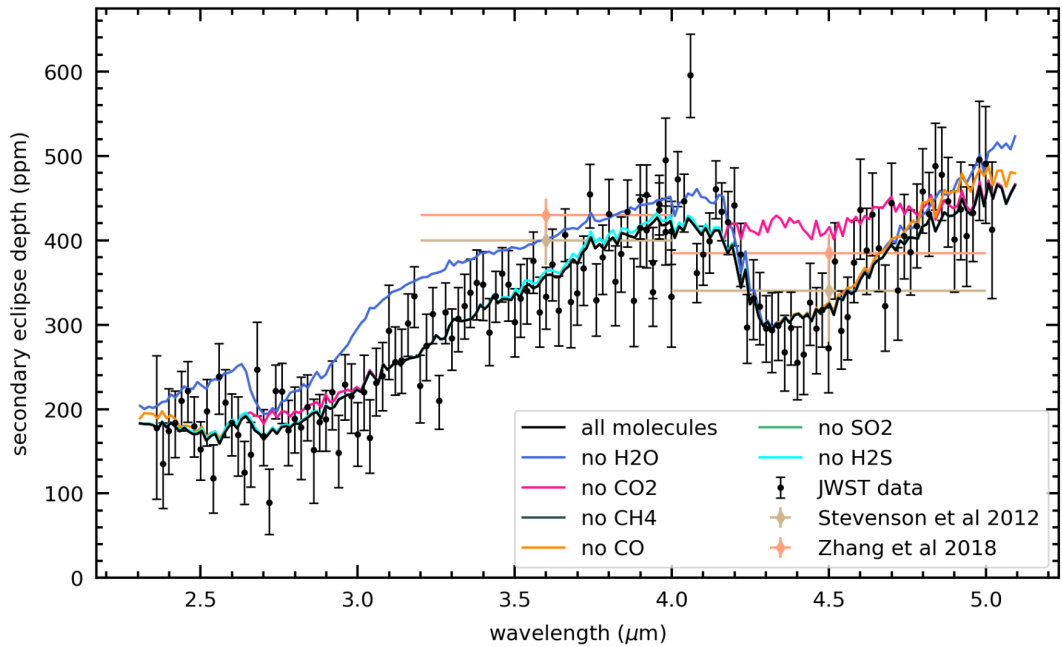

**Figure 1: The thermal emission spectrum of HD 149026b.** The JWST data with 1σ errors are shown as the black circles. Previous data from the Spitzer Space Telescope are shown as diamonds where the horizontal lines indicate the full width of the photometric bandpasses. The best-fit model from the retrieval applied to the JWST data is shown as the black line. This model has [M/H] = 2.37, C/O = 0.81, dilution factor 0.98, and thermal structure as shown in Figure 2. The same model but with the opacities of individual molecules removed one at a time are shown as the colored lines as indicated in the legend. Error bars are 1σ uncertainties.

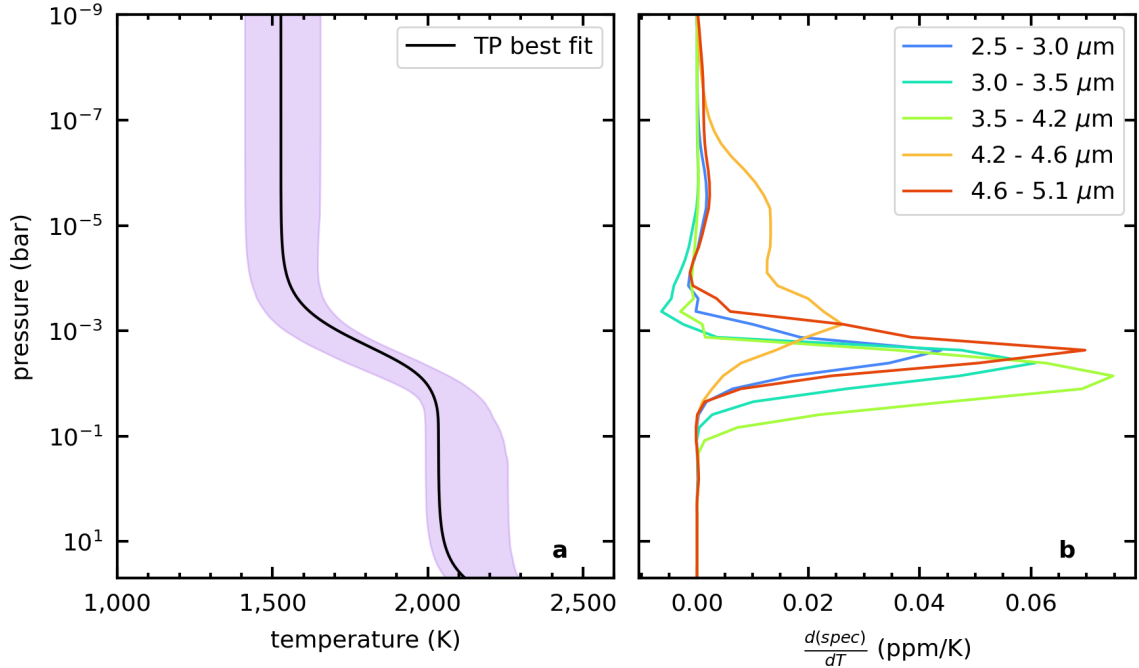

**Figure 2: The T-P profile and contribution functions for HD 149026b.** (a) The best-fit T-P profile and 2$\sigma$ confidence interval from the retrieval samples. (b) The contribution functions for various wavelength ranges in the data are the discrete derivatives of the thermal emission spectrum with respect to small changes in temperature at a given pressure level. Outside the range where there is significant contribution to the spectrum ($10^{-7}$ - $10^{-1}$ bar) the T-P profile morphology is fully predicted by the physically motivated priors. See the Methods for further details.

The T-P profile and contribution functions derived from the retrieval are shown in Figure 2. The presence of spectral absorption features indicates that the temperature decreases with altitude in the infrared photosphere of HD 149026b. That is, there is no thermal inversion. This matches the expectations from theoretical models and empirical trends[27,28], which show that thermal inversions only impact the spectra of planets with dayside temperatures greater than 2,000 K. The contribution functions indicate that the data probe a broad range of pressures from $10^{-7}$ - $10^{-1}$ bar, with the shallowest atmospheric layers being probed by the strong $CO_2$ absorption at 4.2 - 4.6 µm, and the deepest layers probed at the opacity minimum around 4.0 µm.

$CO_2$ is known to be an atmospheric metallicity indicator in giant planet atmospheres[29-31]. Therefore, as expected from the presence of a strong $CO_2$ absorption feature in the spectrum of HD 149026b, the PLATON retrieval favors an elevated metallicity of [M/H] = $2.09^{+0.35}_{-0.32}$ (59 - 275 times solar). The retrieval also favors a carbon-to-oxygen ratio of C/O = 0.84±0.03. The constraint on the C/O is driven by a combination of the detected $H_2O$ and $CO_2$ features, and the lack of detection of other molecules like HCN that would be present in atmospheres with C/O values greater than unity. Importantly, neither the metallicity nor the C/O are strongly correlated with the parameters describing the T-P profile or the dilution factor (see Extended Data Figure 8).

Motivated by the trend that is observed in the solar system, a number of previous studies have attempted to find a correlation between atmospheric metallicity and planet mass for exoplanets (e.g., Refs. [32-34]). However, most of these efforts were hampered by poor chemical abundance constraints due to the limitations of previous data. For example, the high metallicity (>100x solar) previously inferred for the low-density, Saturn-mass planet WASP-39b by Ref. [33] was disputed by other analyses[34-38], and then conclusively revised downward to ~10x solar (i.e., Saturn-like) based on JWST data[16,19,24-26,39].

Figure 3 shows our atmospheric metallicity constraint for HD 149026b in the context of the masses and bulk metallicities of the solar system planets and exoplanets with atmospheric metallicities determined from the spectroscopic detection of both oxygen- and carbon-bearing species. We find that HD 149026b is highly inconsistent with the mass-metallicity trend observed in the solar system. Instead, our result is somewhat more consistent (at 2.1σ) with the trend of atmospheric and bulk metallicity seen in the solar system giant planets. The formation of HD 149026b has been a mystery since it was discovered (e.g. Refs. [41-43]). Our result indicates that the mechanism that led to its bulk metal enrichment similarly enriched its atmosphere.

The atmospheric metallicity of HD 149026b is expected to be 280±26 times solar if the planet is well mixed, whereas the atmospheric metallicity is expected to be ~50 times solar if the planet has a similar atmospheric-to-bulk metallicity fraction as Jupiter and Saturn (~20%, Ref. [12]). Unfortunately, the atmospheric metallicity constraint is not precise enough to distinguish between fully-mixed and layered scenarios for HD 149026b's structure, as both fully mixed and Jupiter/Saturn-like segregation are consistent with the one sigma confidence interval on the measurement. Nevertheless, the correspondence between the bulk and atmospheric metallicity seen in Figure 3 suggests that the elevated C/O ratio of the atmosphere derived here is representative of the bulk envelope. A C/O elevated relative to the solar value (0.59±0.08, Ref. [44]) suggests either a carbon-rich planet forming disk or processes leading to preferential accretion of carbon-rich material (reviewed in Ref. [45]).

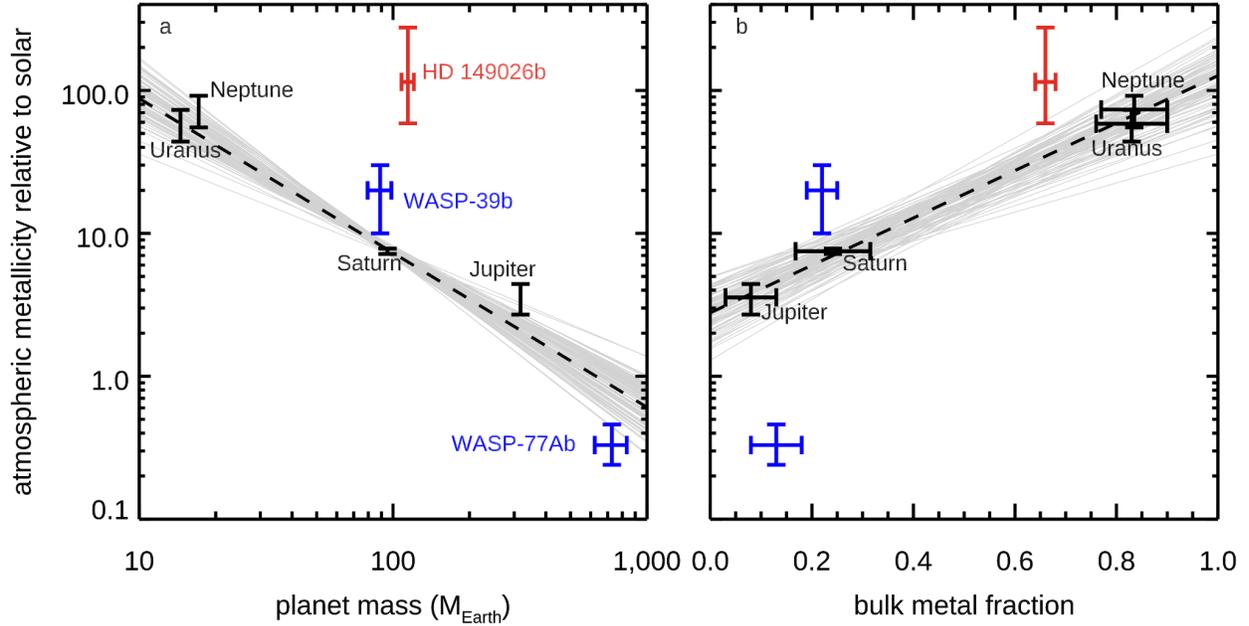

**Figure 3: The metallicity of HD 149026b in context.** Bulk metal fraction (Z) is the mass in metals relative to the total planet mass, and atmospheric metallicity is the number of metal atoms relative to hydrogen atoms normalized to the solar value. The data for the solar system planets are taken from Ref. [5], with the carbon abundance adopted as a proxy for the overall metallicity. The data for WASP-77Ab and WASP-39b are taken from Refs [40] and [39], respectively. The dashed black lines are fits to just the solar system planets, and the solid gray lines are Monte Carlo fits to simulated data points perturbed according to their uncertainties. The fit of the atmospheric metallicity as a function of bulk metal fraction (panel b) has the functional form $A \times 10^{BZ}$, where $A = 2.78 \pm 0.64$ and $B = 1.66 \pm 0.21$. The *p* value of this fit compared to the null hypothesis of no trend is 0.003. Error bars are 1σ uncertainties.

# Methods

**Stellar and planet properties**

As an independent determination of the basic stellar parameters, we performed an analysis of the broadband spectral energy distribution (SED) of the star together with the Gaia DR3 parallax (with no systematic offset applied; see, e.g., Ref. [46]), in order to determine an empirical measurement of the stellar radius, following the procedures described in Refs. [47-49]. We pulled the $B_T V_T$ magnitudes from Tycho-2, the *ubvy* Stromgren magnitudes from Ref. [50], the $JHK_S$ magnitudes from 2MASS, the W1-W4 magnitudes from WISE, and the FUV and NUV magnitudes from GALEX. Together, the available photometry spans the full stellar SED over the wavelength range 0.2 - 22 µm (see Extended Data Figure 1).

We performed a fit using Kurucz stellar atmosphere models, with the free parameters being the effective temperature ($T_{\rm eff}$), surface gravity (log $g$), and metallicity ([Fe/H]). We adopted $T_{\rm eff}$ and log $g$ from the spectroscopic analysis of Ref. [51]; the metallicity we left as a fitted parameter because of the strong constraint provided by the GALEX NUV flux (see below). The remaining free parameter is the extinction $A_V$, which we limited to the maximum line-of-sight extinction from the Galactic dust maps of Ref. [52]. The resulting fit (Extended Data Figure 1) has a reduced $\chi^2$ of 1.2, excluding the GALEX FUV flux which indicates a modest level of chromospheric activity (see below). The best-fit parameters are $A_V$ = 0.00±0.01, $T_{\rm eff}$ = 6085±100 K, and [Fe/H] = 0.25±0.10 (strongly constrained by the GALEX NUV flux, and consistent with the spectroscopically estimated value). Integrating the (unreddened) model SED gives the bolometric flux at Earth, $F_{\rm bol}$ = 1.438±0.017 × $10^{-8}$ erg s$^{-1}$ cm$^{-2}$. Taking the $F_{\rm bol}$ and the Gaia parallax gives the bolometric luminosity directly, $L_{\rm bol}$ = 2.603±0.031 L$_{\rm sun}$. The $F_{\rm bol}$ and $T_{\rm eff}$ together with the Gaia parallax gives the stellar radius, $R_{star}$ = 1.454±0.048 R$_{\rm sun}$. In addition, we can estimate the stellar mass from the empirical relations of Ref. [53], giving $M$ = 1.28±0.08 M$_{\rm sun}$.

The stellar mass and radius derived from the preceding analysis give a stellar density $\varrho$ = 0.586±0.068 g cm$^{-3}$. This is essentially identical to the stellar density implied by the transit parameters derived by Ref. [18] (0.593±0.112 g cm$^{-3}$), but somewhat higher, although still consistent within the 1$\sigma$ errors, than that implied by the transit parameters of Ref. [9] (0.497$^{+0.042}$ $_{-0.057}$ g cm$^{-3}$). HD 149026b has a transit depth of only ~0.3%, which makes it difficult to accurately determine the transit ingress/egress and duration lengths needed to constrain the stellar density from light curves. Therefore, we adopt the stellar mass and radius from the empirical relations and SED fitting. Assuming a stellar radial velocity semi-amplitude of 43.3±1.3 m s$^{-1}$ (Sato et al. 2005) and planet-to-star radius ratio $R_p/R_s$ = 0.0511±0.0011 (taking the average and standard deviation of the values at 3.6 μm and 4.5 μm from Ref. [18]) we derive the mass of the planet $M_p$ = 0.358±0.018 $M_{Jup}$ and the radius of the planet $R_p$ = 0.723±0.029 $R_{Jup}$. Extended Data Figure 2 shows a mass-radius diagram for HD 149026b compared to other transiting planets.

Finally, we can use the star's FUV excess (Extended Data Figure 1) to estimate an age via empirical activity-age relations. The observed FUV excess implies a chromospheric activity of log R'$_{HK}$ = -4.85±0.09 via the empirical relations of Ref. [54]. This estimated activity implies an age of 4.0±1.3 Gyr via the empirical relations of Ref. [55]. As a consistency check, the same empirical relations of Ref. [54] predict from the R'$_{HK}$ a stellar rotation period of 12.3±1.0 days, which is consistent with the (projected) rotation period inferred from the spectroscopically determined $v \sin i$ and the stellar radius determined above, which predicts 14.3±1.4 days.

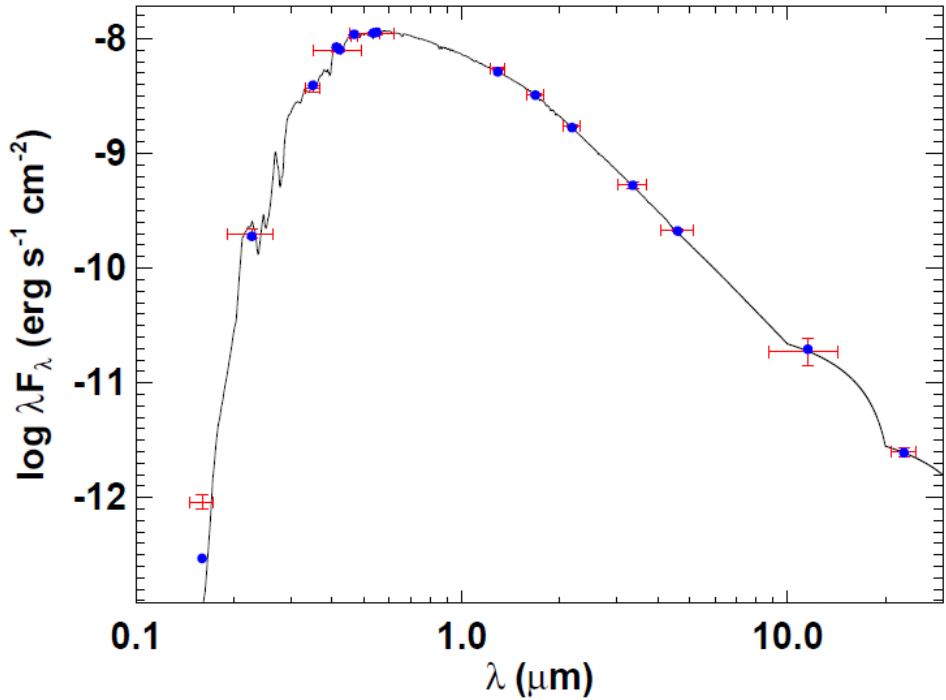

**Extended Data Figure 1: Spectral energy distribution (SED) of HD 149026.** Red symbols represent the observed photometric measurements, where the horizontal bars represent the effective width of the passband. Blue symbols are the model fluxes from the best-fit Kurucz atmosphere model (black). Error bars in flux are 1σ uncertainties; error bars in wavelength give the bandpass of the measurement.

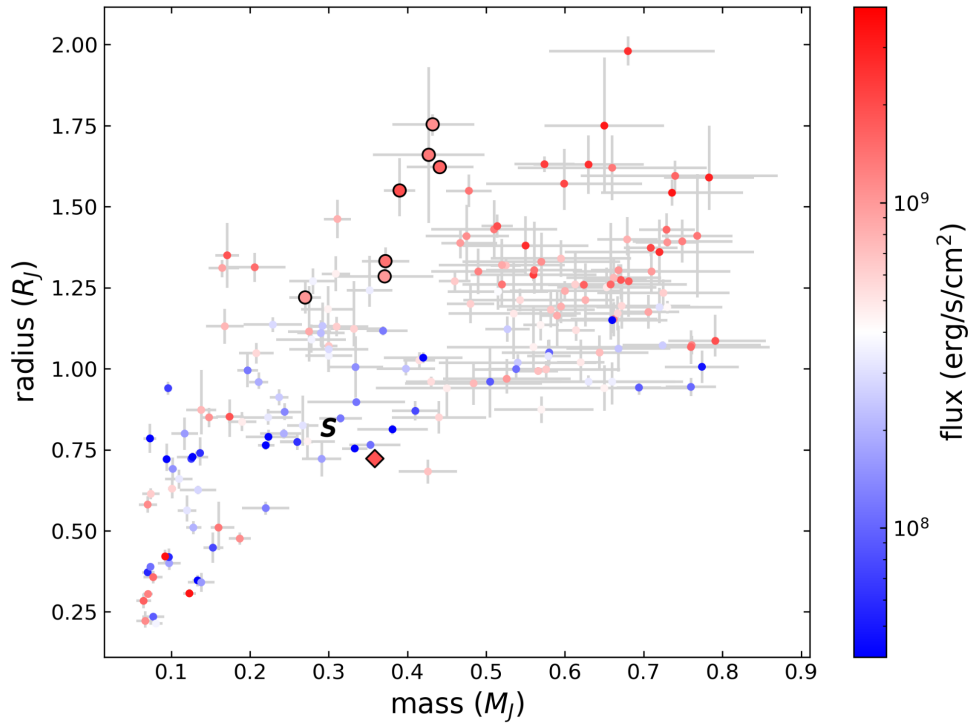

**Extended Data Figure 2: Mass - radius diagram for giant planets like HD 149026b.** The points that are bolded represent planets that are within 20% of HD 149026b's mass and 50% of its irradiation. Saturn's position in the diagram is indicated by the "S". HD 149026b is indicated by the diamond-shaped point. It stands out as being among the smallest planets for its mass despite being highly irradiated. The most similar planet is K2-60b (the point to the right and slightly down from HD 149026b), but it receives just one third of the flux that HD 149026b does. HD 149026b is smaller than Saturn despite receiving >$10^5$ times more irradiation. Error bars are 1σ uncertainties.

**Modeling the interior structure and evolution of the planet**
The bulk composition of HD 149026b was calculated using the interior structure evolution models of Ref. [11]. These models solve the equations of hydrostatic equilibrium, conservation of mass, and an appropriate equation of state to calculate the evolution of a giant planet (especially the radius) with time. The hot Jupiter anomalous heating was set using the scaling relation with flux that was derived empirically in Ref. [11]. The main difference in modeling with that previous work is that we have used a more up-to-date EOS for hydrogen and helium from Ref. [55]. To match our models to the observed parameters (mass, radius, age, and flux) of HD 149026b, we use a Bayesian statistical model described in Ref. [12]. The resulting MCMC retrieval constrains the bulk metallicity of the planet, which we found to be quite high at $Z = .66$ +/- .02 (see Extended Data Figure 3).

With the mass and radius of HD 149026b being so well constrained, modeling uncertainties become an important consideration. The most important for this planet is how to handle the hot Jupiter anomalous heating effect. We have chosen to use the heating power determined from the population[11], but the planet is so small that one might try to argue that anomalous heating is absent here. It turns out this is not sufficient to explain the planet's size – even with no anomalous heating in our model, the planet still requires $Z = .49$ +/- .04. This is a qualitatively identical result: either way the planet requires large amounts of metal to explain its radius, so the most parsimonious theory is that its heating is similar to other hot giants.

More typical sources of uncertainty are the choice of equation of state, the rock-to-ice ratio of the metals, and the core fraction of the planet. Switching back to the older EOS for hydrogen from Ref. [57] only modestly reduces the predicted metallicity ($Z \sim .61$). We assumed a 50-50 rock-to-ice ratio, following the rough guideline of icy bodies in the solar system, but if we instead model all of the metal as rock, we slightly reduce the required metal ($Z \sim .65$). Neither of these changes yields a qualitative difference, but the core fraction is more important. We chose to model a fully-mixed planet, but if we condense some of this metal into a core, significantly more total metal is required to match the radius. This is because a metal-poor envelope is much more thermally expansive, reacting very strongly to the anomalous heating. The most extreme case of a metal-free envelope requires a core fraction as high as .89. This is, however, inconsistent with the high observed atmosphere metallicity. It is also difficult to imagine how a planet would form with such a massive core but without enriching the envelope (metals readily dissolve in liquid metallic hydrogen, e.g. Ref. [58]). Finally, the presence of a core only enhances our qualitative result of an extremely metal-rich planet. As such, we find the conclusion that HD 149026b is very metal-rich to be inescapable.

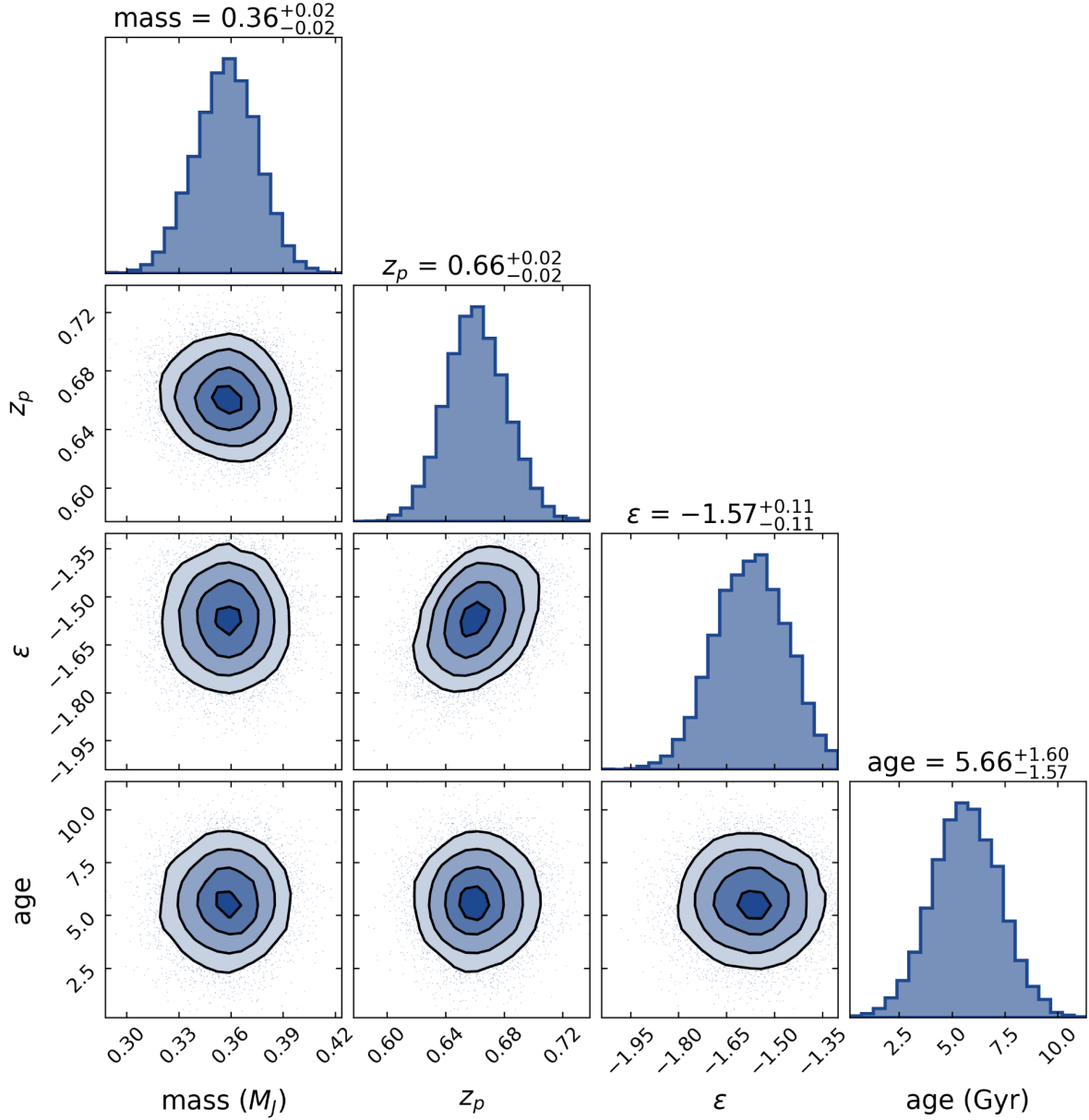

**Extended Data Figure 3: Results of the Bayesian retrieval for the interior structure of HD 149026b.** Mass is in Jupiter masses and age is in gigayears. $Z_p$ and $\varepsilon$ are unitless and refer to the bulk metallicity of the planet and the $\log_{10}$ of the hot Jupiter heating relative to the incident flux, respectively. The retrieval shows that the planet is very metal rich irrespective of the exact age or heating rate. The structure models are from Ref. [11] and the statistical model is from Ref. [12]. Note that the planet radius (accounting for uncertainty) was a vital observation which the model seeks to explain, but it is not a model parameter and so is not shown in the posterior.

**Data reduction and analysis**

**Long wavelength**

We started the data reduction using the uncal.fits files retrieved from the Mikulski Archive for Space Telescopes (MAST) website. The steps in Stage 1 and Stage 2 are identical with the JWST Science Calibration Pipeline (jwst), except that we increased the default jump detection threshold from 4$\sigma$ to 8$\sigma$/6$\sigma$ for F322W2/F444W, and skipped the PhotomStep in jwst Stage 2 because the needed reference files are still a work in progress and this step can introduce noise with pointing jitter.

We then used calints.fits from the Stage 2 outputs to perform background subtraction and spectral extraction. After selecting the subarray of interest, the spectral trace is fit to a gaussian and the center of each column is brought to the center of the detector to perform extraction on a straightened trace. A column-by-column linear fit is performed to calculate the background in the region beyond 11 pixels relative to the center of spectral trace with an outlier rejection threshold = 4$\sigma$/3$\sigma$ for F322W2/F444W. An optimal spectral extraction as defined in Ref. [59] is performed for each integration within 6 pixels on either side of the trace with outlier rejection threshold = 15$\sigma$/10$\sigma$ for F322W2/F444W. This process is illustrated in Extended Data Figure 4. We tested variations of threshold and aperture size in this stage to minimize the median absolute deviation of the final light curves. At the end of this stage, a time-series of 1D spectra is obtained.

We normalized light curve from each detector column and performed a 3$\sigma$/4$\sigma$ outlier rejection on the light curve for F322W2/F444W. We binned the data into 20nm-wide spectroscopic channels and performed outlier rejection again. Following Ref. [16], we only extracted the 2.349 – 4.009 um region for F322W2 and the 3.889 – 5.029 um region for F444W.

We fit the white and spectroscopic light curves with a joint secondary eclipse model that combines the batman code Ref. [60] and a systematics model. The time of the secondary eclipse was determined by fitting the white light curve and then was fixed in spectroscopic channels. The systematics model that we adopted was a product of exponential and linear functions with respect to time of the form: $(c_0 + c_1 t) \cdot (1 + r_0 e^{-t/r_1})$. The size and timescale of the exponential ramp varies between the channels, and it is not even apparent by eye in some channels after trimming the first 30 minutes of data. Therefore, we adopted a uniform prior on the exponential timescale $r_1$ of 5 - 140 minutes to avoid degeneracy between the timescale and amplitude in the channels where the systematic is small. The upper limit on $r_1$ is equal to the amount of observing time before the eclipse after trimming the first 30 minutes of data.

We also tested using a 2nd order polynomial function in time for the systematics model instead of the exponential times linear functions. We found that this form could not reproduce the average morphology seen in the light curves even when trimming up to the first 72 minutes of the data (72 minutes is the longest timescale for the exponential ramp seen in the light curves).

Analyses using the 2nd order polynomial also yielded spectra that were significantly lower overall than the previous Spitzer measurements and that gave reduced $\chi^2$ values in the modeling significantly larger than those found for main result. Therefore, we ultimately discarded this model for the systematics.

All fits were done by Markov Chain Monte Carlo (MCMC) algorithms with the emcee package[61]. We introduced a multiplier to the expected photon noise and fit it for each channel. We found that 14 out of the 140 channels had this multiplier larger than 1.4 (maximum value was 1.98). The final fitted white light curve is shown in Extended Data Figure 5. The fitting results for secondary eclipse time is $2459775.7413\pm0.0008/2459795.8734\pm0.0006$ $BJD_{TDB}$ for F322W2/F444W. In addition, we don't find strong time-correlated noise in the spectroscopic channels, as shown in the Allan deviation plot in Extended Data Figure 6.

**Short wavelength**
The Stage 1 and Stage 2 steps for the short wavelength reduction are identical to those in the long wavelength data reduction with the exception of increasing the jump detection threshold to $10.0\sigma$ for F444W. We then performed an outlier rejection with $7\sigma$ threshold and corrected 1/f noise. We used aperture radii of 55/40 pixels relative to the centroid to extract photometric light curves, and an annulus from 60 to 90 pixels to estimate the background for F322W2/F444W. A $3\sigma$ outlier rejection was then performed on the normalized photometric light curve. We used the same combination of eclipse and systematics model as for the long wavelength data. To analyze the time-correlated noise in the short wavelength data, we used a residual permutation technique (or so called "prayer bead") as defined in Ref. [62]. As can be seen in Extended Data Figure 7, we found significant red noise in the short wavelength data for the second visit.

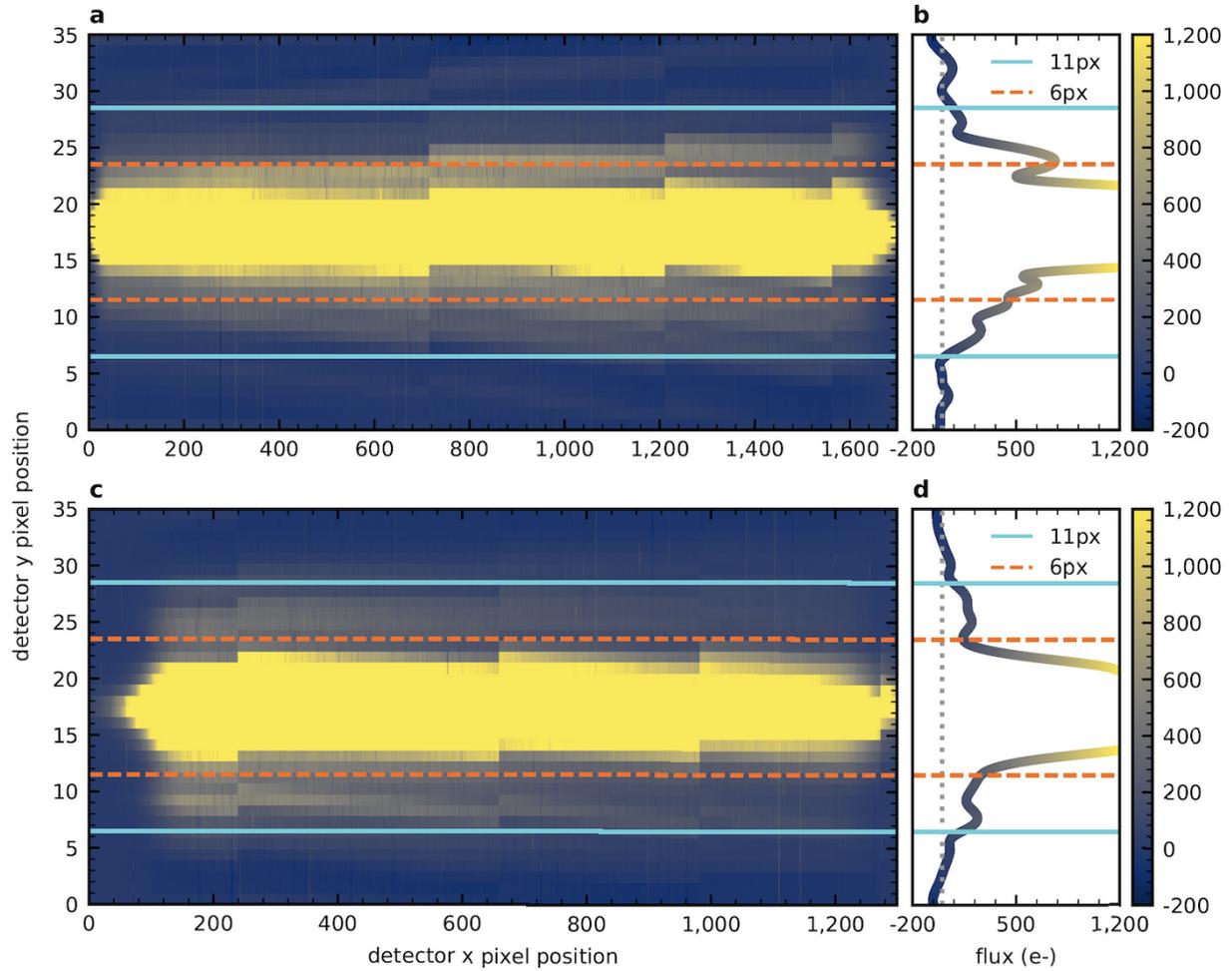

**Extended Data Figure 4: Extraction of the long-wavelength data.** Panels a and c show the spectral trace of the median frame after correcting curvature and subtracting background. Optimal spectral extraction is performed between the orange-dashed lines, and background subtraction is performed in the region above and below the blue lines. Panels b and d show the interpolated cubic function of flux along detector y axis over the 850th to 860th detector columns. Panels a and b show the F322W2 data. Panels c and d show the F444W data.

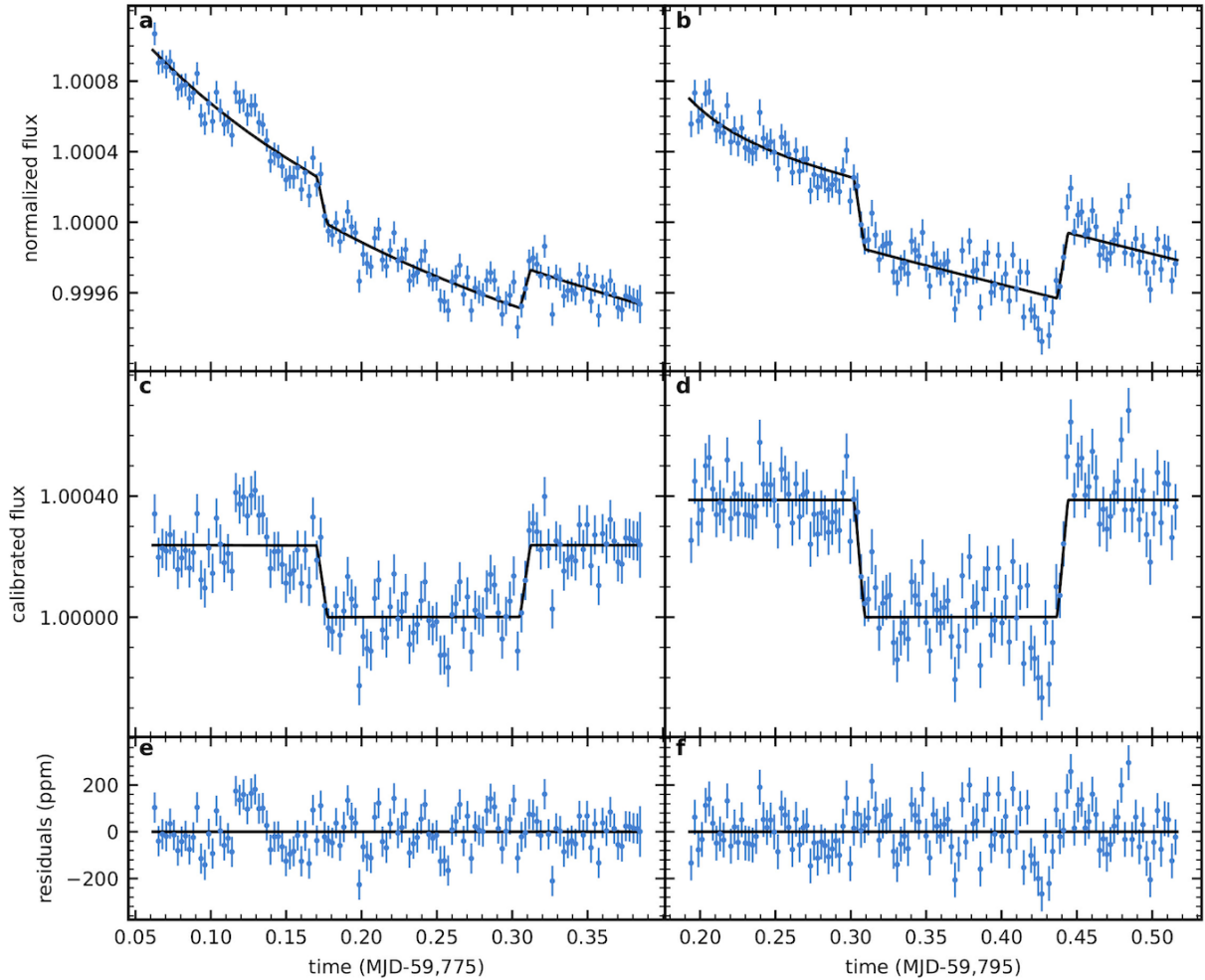

**Extended Data Figure 5: White light curves for the long-wavelength data.** Panels a and b show the data with no corrections and overplotted by best-fit model. Panels c and d show the normalized data with systematics divided out and overplotted by the transit model. Panels e and f show the residuals to the fit. Panels a, c, and e show the F322W2 data. Panels b, d, and f show the F444W data. Error bars are 1σ uncertainties.

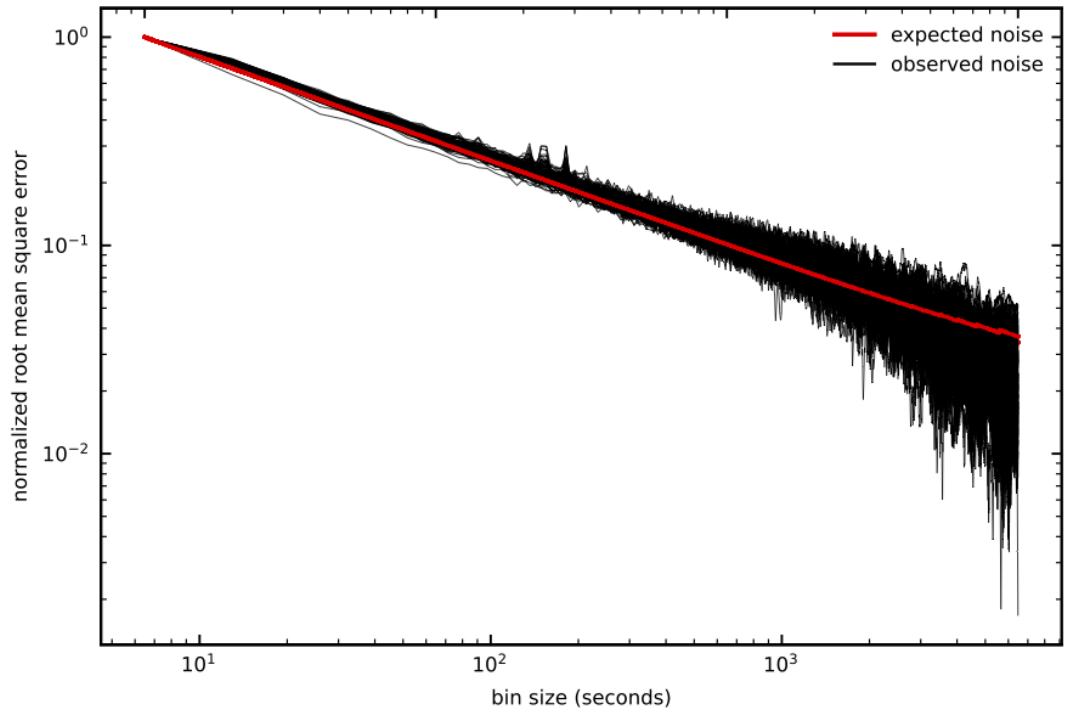

**Extended Data Figure 6: Allan deviation for the spectroscopic light curves obtained in both visits.** The black lines show the root mean square error from each channel, which are normalized by the value of the unbinned data. The red line shows the expected behavior for white noise.

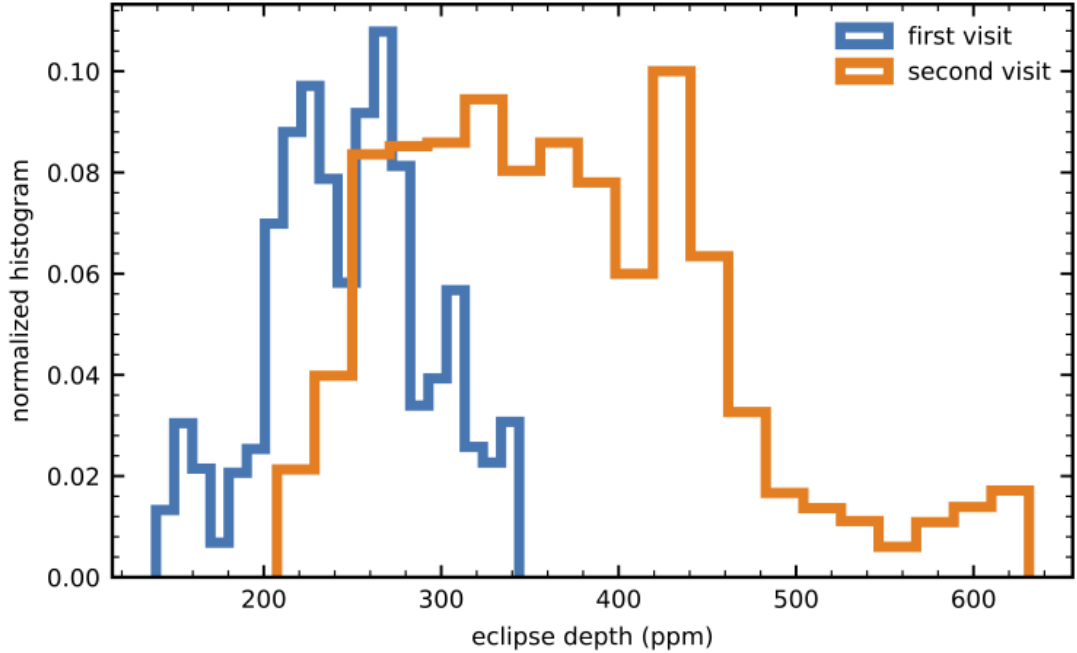

**Extended Data Figure 7: Error analysis for the short-wavelength data.** We compute the residuals between the best-fitting model and the light curve, shift the residuals one point at a time, add the residuals back to the model (in order to preserve the time-correlated structure in the data), and find the best fit again (i.e., a "prayer bead" analysis). The histogram shows the distribution of the eclipse depths for the two visits. The results from the first visit (blue distribution) match the expectation for white noise, but the results for the second visit (green distribution) are highly non-Gaussian, indicating correlated noise in the data.

**Atmospheric retrieval**

We used the PLATON code to model the measured thermal emission spectrum of HD 149026b. The opacity files used for the retrievals are computed at a spectral resolution R=10,000, and the number of live points was progressively increased up to nlive = 1,000 in compliance with the staged approach suggested by Ref. [21]. The estimated atmospheric characteristics and the reduced chi-square only vary slightly through this process, showing robustness in the obtained results.

We assumed the following input parameters from the stellar and planet characterization described above: stellar radius $R_{star}$ = 1.454 $R_{sun}$, stellar temperature $T_{eff}$ = 6085 K, planet radius $R_p$ = 0.723 $R_{Jup}$, and planet mass $M_p$ = 0.358 $M_{Jup}$.. We included additional parameters for the dilution of the dayside spectrum, and we experimented with including a variable $SO_2$ abundance between $10^{-5}$ - $10^{-3}$ bars as motivated by the photochemistry calculations (not shown). The results of the main retrieval without $SO_2$ are summarized in Extended Data Figure 8.

The best-fit model to the data has a reduced $\chi^2 = 0.94$. The dilution factor is consistent with unity (no detected inhomogeneity), indicating that it isn't required to fit these data. There is a weak correlation between the M/H and C/O parameters and the dilution factor, but it doesn't impact their $1\sigma$ confidence intervals. We chose to keep this parameter in the retrieval to marginalize over our ignorance in the degree of inhomogeneity of the dayside of the planet.

**PLATON** allows for a prediction of the atmospheric T-P profile based on the 5-parameter model by Ref. [22] (see Figure 3), as well as chemical abundance profiles of different species using chemical equilibrium computations (see Extended Data Figure 9). In order to understand how much of this information is directly inferred from the thermal emission spectrum, we computed Jacobians for the temperature (as seen in Figure 3) as well as for different species' chemical abundances as shown in Extended Data Figure 10.

The chemistry Jacobians show that the layers probed by the data range in pressures from $10^{-7}$ - $10^{-1}$ bar. Additionally, the spectrum is particularly sensitive to $CO_2$ in the 4.2 - 4.6 μm channels, which correspond to our $CO_2$ detection in the spectrum. It is overall less precise for the $H_2O$ abundance, however practically all channels are sensitive to it. $CH_4$ absorbs in our bandpass, but it is expected to have a very low abundance for our retrieved parameters (see Extended Data Figure 9). Therefore small changes to this baseline abundance would not yield changes to the spectrum. Nevertheless, we would have detected it if for example the atmosphere had C/O > 1.

We note that the uncertainty envelope for the retrieved T-P profile for HD 149026b crosses the condensation curves for MgSi-bearing clouds[63]. Therefore it is possible that ~25% of the oxygen could be sequestered in clouds. This is unlikely to impact the interpretation of the metallicity given the large errors. The impact on the C/O could be larger, but it is likely to be difficult to discern at the individual planet level, and so we leave the study of this for future work.

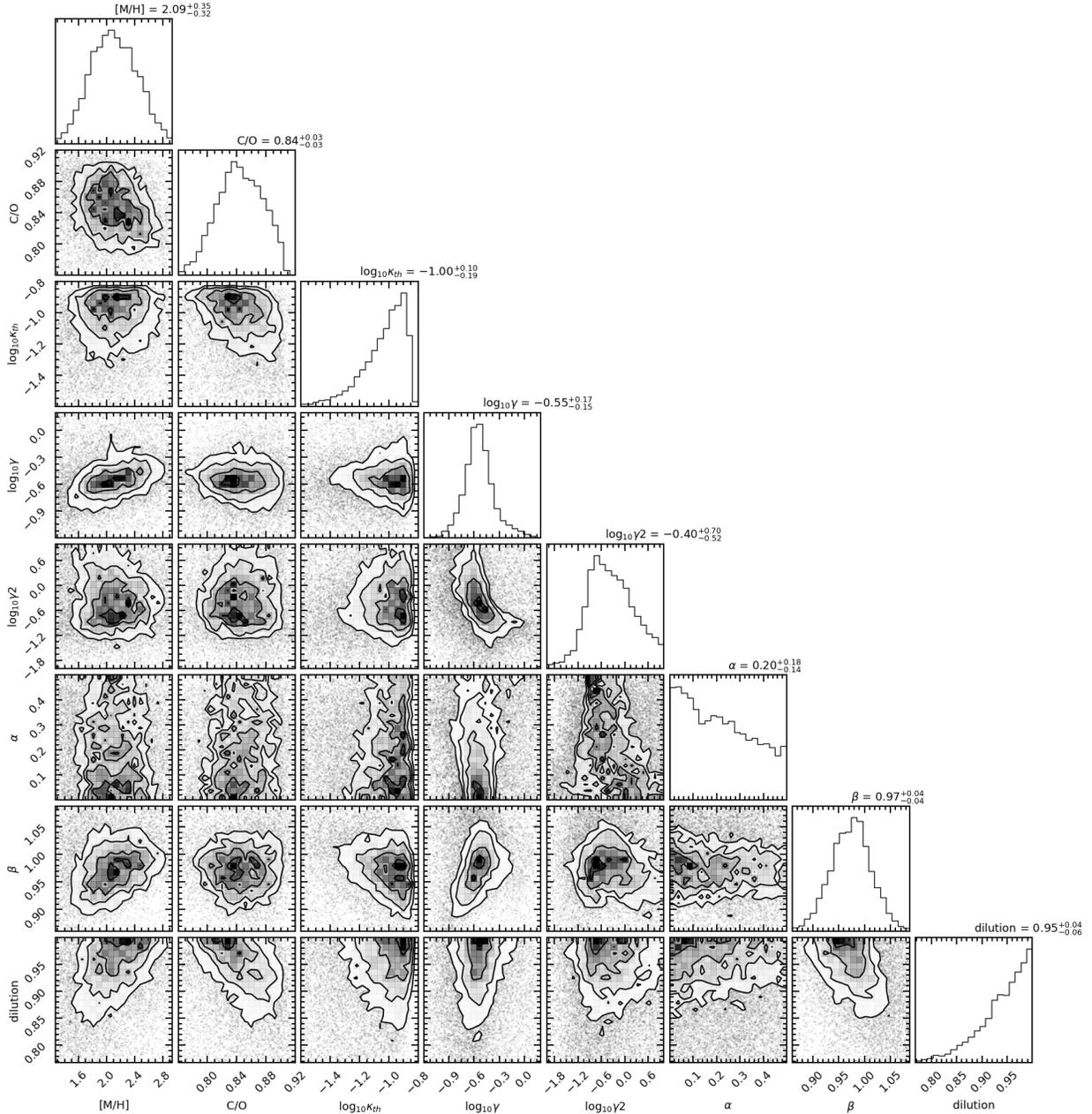

**Extended Data Figure 8: Corner plot of the retrieved parameters for HD 149026b's atmosphere.** From left to right : the metallicity [M/H], the C/O ratio, the 5-parameter T-P profile model from Ref. [22] (thermal opacity, visible to thermal opacity ratio of the first and second visible streams, percentage apportioned to the second visible stream, and effective albedo), and the dilution coefficient as described by Ref. [23].

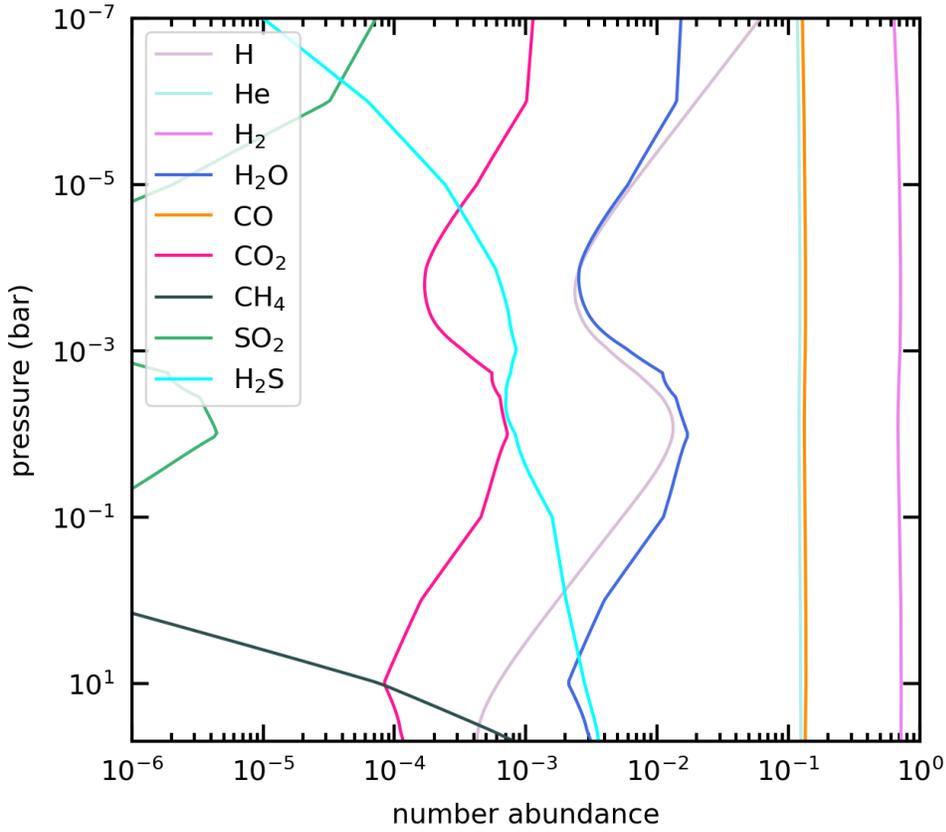

**Extended Data Figure 9: Abundances of key chemical species retrieved for HD 149026b's atmosphere.** These abundances are for the best-fit chemical equilibrium model shown in Figure 2.

**Extended**

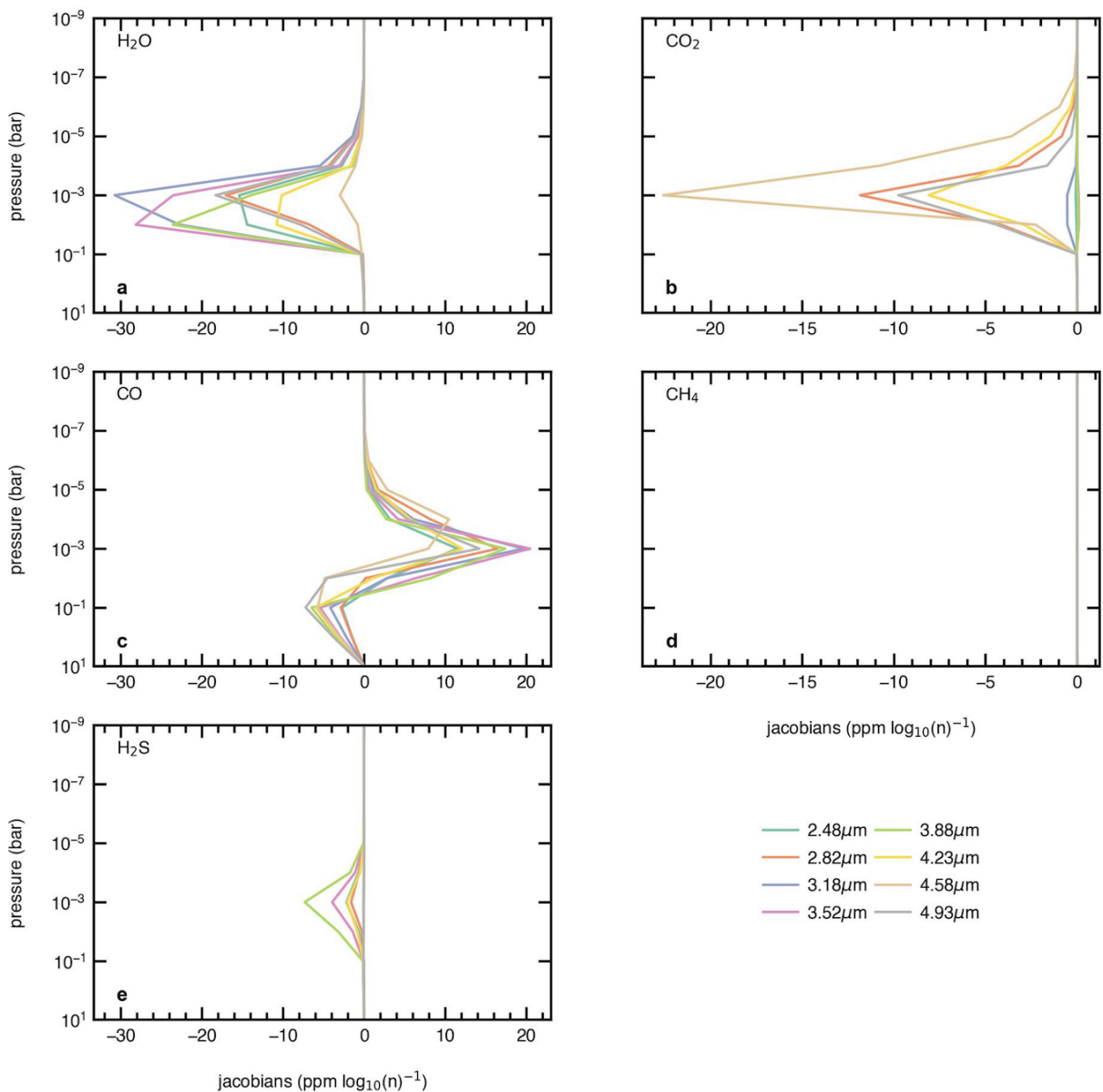

**Data Figure 10: Chemistry Jacobians for HD 149026b.** The spectrum's sensitivity to key molecules is evaluated by computing its deviations from the main fit with respect to small changes in the chemical abundance (in log units). The different coloured lines correspond to different wavelength channels, as shown in the legend.

**Methods References**

**Data Availability**

The data used in this paper are associated with JWST program GTO 1274 (observation numbers 4 and 5) and are available from the Mikulski Archive for Space Telescopes (https://mast.stsci.edu). Science data processing version (SDP_VER) 2022_2a generated the uncalibrated data that we downloaded from MAST. We used JWST calibration software version (CAL_VER) 1.6.0 with modifications described in the text. We used calibration reference data from context (CRDS_CTX) 1004.

**Code Availability**

The Eureka! (https://github.com/kevin218/Eureka) and PLATON (https://github.com/ideasrule/platon) codes used in this study are publicly available.

**Acknowledgements** This work is based on observations made with the NASA/ESA/CSA James Webb Space Telescope. The data were obtained from the Mikulski Archive for Space Telescopes at the Space Telescope Science Institute, which is operated by the Association of Universities for Research in Astronomy, Inc., under NASA contract NAS 5-03127 for JWST. We thank Kevin Stevenson for help with the data reduction.

**Author contributions** J.L.B. led the project and wrote the manuscript. Q.X. performed the data analysis with assistance from M.Z, E.S., E.-M.A., and M.M. P.C.A. performed the atmospheric modeling with assistance from M.Z. and J.I. J.L. provided the JWST data and contributed text. D.T. performed the interior structure modeling. S.-M.T. did the photochemistry calculations. K.G.S. performed the analysis of the host star properties. All authors commented on the manuscript and aided in the interpretation.

**Competing interests** The authors declare no competing interests.

**Additional information**
**Supplementary information** none
**Correspondence and requests for materials** should be addressed to Jacob Bean (jbean@astro.uchicago.edu).